\documentclass[journal]{IEEEtran}

\normalsize

\usepackage[margin=1.00in]{geometry} 
\usepackage{ifpdf}
 \ifpdf
     \usepackage[pdftex]{graphicx}
 \else
     \usepackage[dvips]{graphicx}
 \fi

\graphicspath{{figures/}}

\usepackage{cite}
\usepackage{subfigure}

\usepackage{epstopdf}					
\usepackage[cmex10]{amsmath}
\usepackage{amssymb}

\usepackage{algorithm}

\usepackage{algorithm}
\usepackage{algpseudocode}
\usepackage{graphicx}
\usepackage{xcolor}
\usepackage{mdwmath}
\usepackage{mdwtab}
\usepackage{flushend}

\begin{document}
%
\title{Second-order Cyclostationarity-based Detection of LTE SC-FDMA Signals for  \\ Cognitive Radio Systems}
%
%
%


\author{Walid~A.~Jerjawi, Yahia~A.~Eldemerdash,~\IEEEmembership{Student Member,~IEEE,}
       and \\ Octavia~A.~Dobre,~\IEEEmembership{Senior Member,~IEEE}
     
\thanks{Walid~A.~Jerjawi, Yahia A. Eldemerdash, and  Octavia A. Dobre are with the Faculty of Engineering and Applied Science, Memorial
University of Newfoundland,  St. John's, Canada. 
Email: \{wj2222, yahia.eldemerdash, odobre\}@mun.ca.}
\vspace{-0.4cm}           
}

\maketitle

\vspace{-2.5cm}
\begin{abstract}
In this paper, we investigate the detection of long term evolution (LTE) single carrier-frequency division multiple access (SC-FDMA) signals, with application to cognitive radio systems. We explore the second-order cyclostationarity of the LTE SC-FDMA signals, and apply results obtained for the cyclic autocorrelation function to signal detection. The proposed detection algorithm provides a very good performance under various channel conditions, with a short observation time and at low signal-to-noise ratios, with reduced complexity. The validity of the proposed algorithm is verified using signals generated and acquired by laboratory instrumentation, and the experimental results show a good match with computer simulation results.

\end{abstract}


\section{Introduction}
 
\IEEEPARstart{W}{ith}
the increasing demand for high data rate services, which require increased bandwidth, the inadequacy of fixed  radio spectrum allocation has become a serious problem. According to the current spectrum assignment policies, a specific band is assigned to a certain wireless system. While this resolves the interference problem between different  systems, it leads to spectrum scarcity. On the other hand, spectrum is available at various times and geographical locations, and the current assignment basically renders its under-utilization. Dynamic spectrum access (DSA) refers to communication techniques that exploit the spectrum holes to increase spectrum utilization. Cognitive radio (CR)  \cite{haykin2005cognitive,yucek2009survey,axell2012spectrum,bao2013histogram,barbe2011automatic} is seen as a potential approach to DSA implementation, which allows the co-existence of  primary/incumbent and  secondary/cognitive users.  For that, the CR needs to be aware of the spectrum environment, which requires the detection/identification of the primary users \cite{haykin2005cognitive,yucek2009survey,axell2012spectrum,bao2013histogram,barbe2011automatic}. Signal cyclostationarity-based methods have been shown to be suitable to performing such a task, with the advantages that they are not sensitive to the noise uncertainty, and require neither channel estimation nor timing and frequency synchronization \cite{haykin2005cognitive,yucek2009survey,dobre2007survey}. The detection/identification of the frequency division multiplexing (OFDM) and single carrier signals was extensively studied in the literature (see, e.g., \cite{bao2013histogram,barbe2011automatic,dobre2007survey,Alla_WiMAX_classification,cl_OFDM1,oner2007extraction,
socheleau2011cognitive,kim2008specific,adrat20092nd,sutton2008cyclostationary,bouzegzi2010new,
grimaldi2007automatic,zhang2013second}). For example, the algorithms in \cite{Alla_WiMAX_classification,cl_OFDM1,oner2007extraction} used the cyclic prefix (CP)-induced cyclic statistics, whereas the pilot-induced cyclic statistics were employed in \cite{socheleau2011cognitive}. For the former, the fact that the CP structure is repeated every OFDM symbols was exploited, while the repetition of the pilot pattern in time and frequency was used in the latter. The pilot pattern was considered for the detection/identification of WiFi, digital video broadcasting-terrestrial, and fixed worldwide interoperability for microwave access (WiMAX) signals. Furthermore, the preamble-induced cyclostationarity was used in \cite{Alla_WiMAX_classification} for the detection/identification of mobile WiMAX signals; this is based on the repetition of the preamble every frame, as well as the existence of a repetitive pattern of subcarriers in the preamble. The second-order cyclostationarity was exploited in \cite{kim2008specific} to identify diverse IEEE 802.11 standard signals, while a theoretical analysis of the second-order cyclostationarity induced by pilots was performed in \cite{adrat20092nd} with application to the IEEE 802.11a signal detection/identification. Furthermore, cyclostationarity signatures intentionally introduced in the signal were exploited in \cite{sutton2008cyclostationary}; these are due to the redundant transmission of message symbols on more than one subcarrier. By using this approach, signals can be uniquely detected by the cycle frequency (CF) created by the embedded signature. Subcarrier mapping permits cyclostationarity signatures to be inserted in the data-carrying waveforms without adding significant complexity to the existing transmitter design; however, it has the disadvantage of resulting in extra overhead. 
On the other hand, the study of  single carrier-frequency division multiple access
(SC-FDMA) signal detection has not been performed yet in the literature; SC-FDMA signals are used as the alternative to OFDM for uplink traffic in long term evolution (LTE) systems \cite{myung2008single}. In this paper, for the first time in the literature, we study the second-order cyclostationarity of LTE SC-FDMA signals, and its application to their detection. We model the signals and derive the analytical closed-form expressions of their cyclic autocorrelation function (CAF) and CFs. Based on these results, we propose an algorithm for the detection of the LTE SC-FDMA signals. The proposed algorithm provides a good detection performance at low signal-to-noise ratios (SNRs), short observation time, and under diverse channel conditions.

The rest of this paper is organized as follows. Section II introduces the signal model and Section III presents the study of SC-FDMA signal second-order cyclostationarity. The proposed algorithm for signal detection is described in Section IV, while simulation and experimental results are discussed in Section V. Conclusions are drawn in Section VI.

\section{Signal Model\label{sec:SIGNAL-MODEL}}

\begin{figure*}
\begin{centering}
\includegraphics[width=1\textwidth]{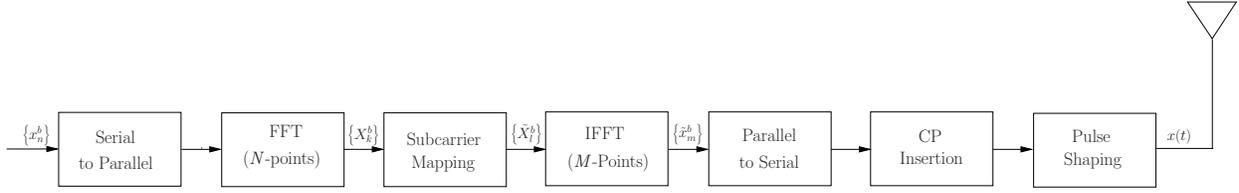}
\par\end{centering}

\caption{SC-FDMA signal generation.\label{fig:signal_generation}}
\end{figure*}

Fig. \ref{fig:signal_generation} shows the  SC-FDMA signal generation scheme. The 
 data symbols of the $b$-th input block $\left\{ x_{n}^{b}\right\} _{n=0}^{N-1}$ correspond either to a phase-shift-keying (PSK) or a quadrature amplitude modulation (QAM) signal constellation. The serially modulated data symbols are then converted into $N$ parallel data streams, and passed through an $N$-point fast
Fourier transform (FFT) block, which generates the frequency domain
symbols $\left\{ X_{k}^{b}\right\} _{k=0}^{N-1}$. Then, the output
of the FFT is passed through the subcarrier mapping block. This assigns
the $\left\{ X_{k}^{b}\right\} _{k=0}^{N-1}$ symbols to $M\geq N$
subcarriers, usually in a localized mode (LFDMA) \cite{myung2008single}. Note
that $M=NQ$, with $Q$ as the expansion factor, and the unoccupied
subcarriers are set to zero. The output symbols in LFDMA, $\{ \tilde{X}_{l}^{b}\} _{l=0}^{M-1}$,
are frequency domain samples; these are passed through an $M$-point
inverse FFT (IFFT), yielding  the time domain samples $\left\{ \tilde{x}_{m}^{b}\right\} _{m=0}^{M-1}$. 
Let $m=Qn+q$, with $ n=0,..., N-1$
and $ q=0,...,Q-1$. Then, the time domain samples can be expressed
as \cite{myung2008single}
\begin{equation}
\tilde{x}_{m}^{b}=\begin{cases}
\begin{array}{l}
\frac{1}{Q}{x^b_{n}},\\
\\\frac{\left(1-e^{j\frac{2\pi q}{Q}}\right)}{QN}\overset{N-1}{\underset{p=0}{\sum}}\frac{x_{p}^{b}}{1-e^{j2\pi\left(\frac{\left(n-p\right)}{N}+\frac{q}{QN}\right)}},\end{array} & \begin{array}{c}
q=0,\\
\\q\neq0.\end{array}\end{cases}\label{eq:1}\end{equation}


\begin{figure*}
\begin{centering}
\includegraphics[width=1\textwidth]{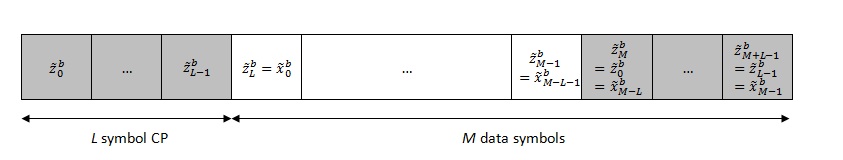}
\par\end{centering}

\caption{SC-FDMA block with CP.\label{fig:tx_block}}
\end{figure*} 

From (\ref{eq:1}), one can see that the time-domain LFDMA signal
samples consist of copies of
the input time symbols scaled by a factor of {1}/${Q}$ for the $N$-multiple sample positions; between these positions, the
signal samples are the weighted sums of the time symbols in the input
block.

To combat the inter-symbol interference caused by the channel delay spread, a cyclic prefix (CP) of $L$ samples is added at the beginning of every $M$ samples $(L<M)$ for each block. According to  the standards \cite{myung2008single,andrews2007fundamentals}, $L$ is considered a multiple integer of $Q$. A root raised cosine (RRC) filter is used to pulse shape the SC-FDMA signal \cite{myung2008single}. The structure of an SC-FDMA block with CP is presented in Fig. \ref{fig:tx_block}, where  $\tilde{z}_{u}^{b}$ represents the symbol transmitted within the $u$-th symbol period of block $b$, $u=0,...,M+L-1$.
The symbol $\tilde{z}_{u}^{b}$ is expressed as

\vspace{-0.2cm}
\begin{equation}
\tilde{z}_{u}^{b}=\begin{cases}
\begin{array}{c}
\tilde{x}_{M-L+u}^{b},\\
\\\tilde{x}_{u-L}^{b},\end{array} & \begin{array}{c}
u=0,...,L-1,\\
\\u=L,...,M+L-1.\end{array}\end{cases}\label{eq:3}\end{equation}

Hence, the received noise-free SC-FDMA signal can be expressed as

\begin{equation}
r(t)=\sum_{b=-\infty}^{\infty}\left[\sum_{u=0}^{M+L-1}\tilde{z}_{u}^{b}g(t-uT-b(M+L)T)\right],\label{eq:2}\end{equation}
where $g(t)$ represents the pulse shape and $T$ is the symbol duration.

\section{Second-order Cyclostationarity of SC-FDMA Signals \label{sec:Second}}
\subsection{Definitions}
A random process $r(t)$ is said to be second-order cyclostationary
if its mean and time-varying autocorrelation function (AF) are almost
periodic functions of time \cite{spooner1994cumulant}. The latter
is expressed as a Fourier series \cite{spooner1994cumulant}

\vspace{-0.3cm}
\begin{equation}
c_{r}(t,\tau)=\textrm{E}[r(t)r^{*}(t-\tau)]=\sum_{\beta\in\kappa}c_{r}(\beta,\tau)e^{j2\pi\beta t},\label{eq:4}\end{equation}
where $\textrm{E}\left[.\right]$ represents the expectation operator,
the superscript {*} denotes complex conjugation, and $c_{r}(\beta,\tau)$
is the CAF at CF $\beta$ and delay $\tau,$ which can be expressed
as \cite{spooner1994cumulant}

\vspace{-0.5cm}
\begin{equation}
c_{r}(\beta,\tau)=\lim_{I\rightarrow\infty}I^{-1}\int_{-I/2}^{I/2}c_{r}(t,\tau)e^{-j2\pi\beta t}\, dt,\label{eq:5} \end{equation}
and $\kappa=\{\beta:\, c_{r}(\beta,\tau)\neq0\}$ represents the set
of CFs.

For a discrete-time signal $r(u)$  obtained by periodically sampling
the continuous-time signal $r(t)$ at rate $f_{s},$ the CAF at
CF $\tilde{\beta}$ and delay $\tilde{\tau}$ is given by%
\footnote{Note that these results are valid under the assumption of no aliasing
in the cycle and spectral frequency domains.%
} \cite{spooner1994cumulant}

\begin{equation}
c_{r}(\tilde{\beta},\tilde{\tau})=c_{r}(\beta f_{s}^{-1},\tau f_{s}), 
\label{eq:6} \end{equation}
where $\tilde{\beta}=\beta f_{s}^{-1}$ and $\tilde{\tau}=\tau f_{s}$.

The estimator of CAF at CF $\tilde{\beta}$ and delay $\tilde{\tau}$,
based on $U_{s}$ samples, is given by \cite{spooner1994cumulant}

\begin{center}
\begin{equation}
\hat{c}_{r}(\tilde{\beta},\tilde{\tau})={U_{s}}^{-1}\sum_{u=0}^{U_{s}-1}r(u)r^{*}(u-\tilde{\tau})e^{-j2\pi\tilde{\beta}u}.\label{eq:7}\end{equation}

\par\end{center}

\subsection{Analytical Expressions of AF, CAF and Set of CFs for the SC-FDMA Signals}
By using  (\ref{eq:2}) and (\ref{eq:4}), the AF of the SC-FDMA signals can be written as \footnote{ Note that these results are for the CAF of the signal component and that of the noise component must be added to the final result.}


\vspace{-0.25cm}
\begin{equation}
\begin{array}{ll}
c_{r}(t,\tau) & =\overset{\infty}{\underset{b_{1}=-\infty}{\sum}}\overset{\infty}{\underset{b_{2}=-\infty}{\sum}}\overset{M+L-1}{\underset{u_{1}=0}{\sum}}\overset{M+L-1}{\underset{u_{2}=0}{\sum}}\textrm{E}\left[\tilde{z}_{u_{1}}^{b_{1}}(\tilde{z}_{u_{2}}^{b_{2}})^{*}\right]\\
& \times g(t-u_{1}T-b_{1}(M+L)T)\\
 & \times g^{*}(t-u_{2}T-b_{2}(M+L)T-\tau).\end{array}\label{eq:8}\end{equation}

%
%
%
%
%
%
%
%
%
%
Further, through cumbersome mathematical calculations, one can show that the analytical closed form expression of the AF is given as:

\begin{subequations}\label{eq:44}
\begin{multline} 
\bullet \textrm{ for }{\tau}=\pm{\tau}_{s},\:0\leq{\tau}_{s}<T, \textcolor{white}{............................................................................................} \\
 c_{r}(t;{\tau})=\left(\frac{T-{\tau}_{s}}{4T}+\frac{{\tau}_{s}}{2NT}\frac{1}{1-e^{j\pi\frac{1}{N}}}\right)c_{x}\left[g(t)g^{*}(t)\right]
\\ 
 \otimes\overset{\infty}{\underset{k=-\infty}{\sum}}\delta(t-kT), \textcolor{white}{...............................} \end{multline} 
 
\begin{multline}
\bullet \textrm{ for }{\tau}={\mu}T+\textrm{sgn}(\mu){\tau}_{s}, \: \mu=\pm1,\pm3, \ldots, \\ \pm(M-L-1), \:0\leq{\tau}_{s}<T, \\
c_{r}(t;{\tau})=\frac{T-{\tau}_{s}}{T}A(\mu)\left[g(t)g^{*}(t)\right]\otimes\overset{\infty}{\underset{k=-\infty}{\sum}}\delta(t-kT), \\
\textrm{ where }      
A(\mu)=\frac{c_{x}}{2N}\frac{1}{1-e^{j\pi\frac{\mid\mu\mid}{N}}}, \textcolor{white}{...............................}
\end{multline} 

\vspace{-0.45cm}
\begin{multline}
\bullet \textrm{ for }{\tau}={\mu}T+\textrm{sgn}(\mu){\tau}_{s}, \: \mu=\pm2,\pm4, \ldots, \\ \pm(M-L),
\:0\leq{\tau}_{s}<T,\\
c_{r}(t;{\tau}) \hspace{-0.05cm} = \hspace{-0.05cm} \frac{{\tau}_{s}}{T}A(\mu+1)\left[g(t)g^{*}(t)\right]\otimes \hspace{-0.2cm} \overset{\infty}{\underset{k=-\infty}{\sum}} \hspace{-0.15cm} \delta(t-kT),\end{multline}

\vspace{-0.45cm} 
\begin{multline}
\bullet \textrm{ for }{\tau}={\mu}T+\textrm{sgn}(\mu){\tau}_{s},\:\mu=\pm M,\:0\leq {\tau}_{s}<T, \\
c_{r}(t;{\tau})=\left(\frac{T-{\tau}_{s}}{4T}+\frac{{\tau}_{s}}{2NT}\frac{1}{1-e^{\pm j\pi\frac{1}{N}}} 
\right) 
\\
\hspace{1cm}\times c_{x}\overset{L-1}{\underset{u=0}{\sum}}\left[g(t-uT)g^{*}(t-uT)\right] \textcolor{white}{..}
\\ \hspace{-4.5cm} \otimes\overset{\infty}{\underset{b=-\infty}{\sum}}\delta(t-b(M+L)T), \textcolor{white}{.........}
\end{multline}

\vspace{-0.45cm}
\begin{multline}
\bullet \textrm{ for }{\tau}={\mu}T+\textrm{sgn}(\mu){\tau}_{s},\:\mu=\pm(M-L+1), \\ \pm(M-L+3), \ldots,
\pm(M+L-1),0\leq {\tau}_{s}<T, 
\\
c_{r}(t;{\tau})=\frac{T-{\tau}_{s}}{T}A(\mu)\overset{L-1}{\underset{u=0}{\sum}}\left[g(t-uT)g^{*}(t-uT)\right]
\\
\otimes\overset{\infty}{\underset{b=-\infty}{\sum}}\delta(t-b(M+L)T), \textcolor{white}{...............}
\end{multline}

\vspace{-0.45cm}
\begin{multline}
\bullet \textrm{ for }{\tau}={\mu}T+\textrm{sgn}(\mu){\tau}_{s},\:\mu=\pm(M-L+2),\ldots,
\\
\pm(M-2),\pm(M+2),\ldots\pm(M+L-2), 0\leq {\tau}_{s}<T,\textcolor{white}{.......................................................................................}\\
c_{r}(t;{\tau})=\frac{{\tau}_{s}}{T}A(\mu+1)\overset{L-1}{\underset{u=0}{\sum}}\left[g(t-uT)g^{*}(t-uT)\right]
\\
\otimes\overset{\infty}{\underset{b=-\infty}{\sum}}\delta(t-b(M+L)T), \textcolor{white}{...............}
\end{multline}

\vspace{-0.85cm}
\begin{multline}
 \bullet \textrm{ otherwise } \\
c_{r}(t;{\tau})=0. \textcolor{white}{................................................}
 \end{multline}
\end{subequations}

In the following, we provide the proof for the zero-delay case for illustration. The same procedure  is applicable to other cases, as one can see from \cite{Jerjawi2014second};  details are not included here due to the space considerations.
%
\vspace{0.3cm}
\begin{IEEEproof}

 From (\ref{eq:8}), one can notice that $c_{r}(t,0)$ has non-zero
significant values when $b_{1}=b_{2}=b$ (the same data block) and
$u_{1}=u_{2}=u$  (the same symbol within a block). Based on this observation,
(\ref{eq:8}) can be expressed as

\begin{equation}
\begin{array}{lll}
c_{r}(t;0)= &\displaystyle \sum_{b=-\infty}^{\infty}\sum_{u=0}^{M+L-1}\textrm{E}\left[\tilde{z}_{u}^{b}(\tilde{z}_{u}^{b})^{*}\right]\\
&\hspace{-0.6cm} \times g(t-uT-b(M+L)T)\\
& \hspace{-0.6cm} \times g^{*}(t-uT-b(M+L)T).
\end{array}\label{eq:9}
\end{equation}

Furthermore, by emphasizing the  summation over the CP symbols, and considering $u$ odd and $u$ even,
(\ref{eq:9}) is  expressed as

\hspace{-0.5cm} \begin{equation}
\begin{array}{ll}

\hspace{-0.4cm}c_{r}(t;0)= & \hspace{-0.5cm}\overset{\infty}{\underset{b=-\infty}{\sum}}(\overset{L-1}{\underset{\underset{\textrm{even}}{u=0}}{\sum}}\textrm{E}\left[\tilde{z}_{u}^{b}\left(\tilde{z}_{u}^{b}\right)^{*}\right] g(t-uT-b(M+L)T)\\
&\hspace{-1.8cm} \times g^{*}(t-uT-b(M+L)T))+\overset{\infty}{\underset{b=-\infty}{\sum}}(\textrm{ }\overset{L-1}{\underset{\underset{\textrm{odd}}{u=0}}{\sum}}\textrm{E}\left[\tilde{z}_{u}^{b}\left(\tilde{z}_{u}^{b}\right)^{*}\right] \\

& \hspace{-1.8cm} \times g(t-uT-b(M+L)T)g^{*}(t-uT-b(M+L)T))\\

 &  \hspace{-1.8cm}  +\overset{\infty}{\underset{b=-\infty}{\sum}}(\textrm{ }\overset{M+L-1}{\underset{\underset{\textrm{even}}{u=L}}{\sum}}\textrm{E}\left[\tilde{z}_{u}^{b}\left(\tilde{z}_{u}^{b}\right)^{*}\right]g(t-uT-b(M+L)T)\\
 &  \hspace{-1.8cm} \times g^{*}(t-uT-b(M+L)T))+\hspace{-0.25cm}\overset{\infty}{\underset{b=-\infty}{\sum}}(\textrm{ } \hspace{-0.2cm} \overset{M+L-1}{\underset{\underset{\textrm{odd}}{u=L}}{\sum}}\textrm{E}\left[\tilde{z}_{u}^{b}\left(\tilde{z}_{u}^{b}\right)^{*}\right]\\
 & \hspace{-1.8cm} \times g(t-uT-b(M+L)T)g^{*}(t-uT-b(M+L)T)).\end{array}\label{eq:11}\end{equation}

Henceforth, we refer to the first, second, third, and fourth terms in the right hand side of  (\ref{eq:11}) as  
$\mathbb{\mathfrak{c}}_{r}^{(1)}(t;0)$, $\mathbb{\mathfrak{c}}_{r}^{(2)}(t;0)$, $\mathbb{\mathfrak{c}}_{r}^{(3)}(t;0)$, and $\mathbb{\mathfrak{c}}_{r}^{(4)}(t;0)$, respectively.  We consider each of these terms, starting with the last two. As such, with $m=u-L$ and using (\ref{eq:3}), $\mathbb{\mathfrak{c}}_{r}^{(3)}(t;0)$ becomes

\begin{equation}
\begin{array}{ll}
\mathbb{\mathfrak{c}}_{r}^{(3)}(t;0)= & \overset{\infty}{\underset{b=-\infty}{\sum}}\overset{M-1}{\underset{\underset{\textrm{even}}{m=0}}{\sum}}\textrm{E}\left[\tilde{x}_{m}^{b}\left(\tilde{x}_{m}^{b}\right)^{*}\right]\\ & 
 \hspace{-0.6cm}\times g(t-(m+L)T-b(M+L)T)\\
 &  \hspace{-0.6cm} \times g^{*}(t-(m+L)T-b(M+L)T).\end{array}\label{eq:12}\end{equation}
  
  By considering that   $m=2n$, $n=0,...,N-1$, and
  by replacing (\ref{eq:1}) into (\ref{eq:12}) with $Q=2$  \footnote{ Note that in the  LTE standard $Q$ is approximately equal to  1.7 \cite{e-utra}. For simplification purposes, here we consider  the closest integer value, i.e., $Q=2$. }, $\mathbb{\mathfrak{c}}_{r}^{(3)}(t;0)$ becomes

\begin{equation}
\begin{array}{l}
\mathbb{\mathfrak{c}}_{r}^{(3)}(t;0) =\frac{c_{x}}{4}\overset{\infty}{\underset{b=-\infty}{\sum}}\overset{M-1}{\underset{\underset{\textrm{even}}{m=0}}{\sum}}g(t-(m+L)T-b(M+L)T)\\ \hspace{1.4cm}\times g^{*}(t-(m+L)T-b(M+L)T),\end{array}\label{eq:13}\end{equation}
where $c_{x}=\textrm{E}[x_{n}^{b}(x_{n}^{b})^{*}]$ represents the
correlation corresponding to the points in the signal constellation.

Similarly, with $m=u-L$ and using (\ref{eq:3}), $\mathbb{\mathfrak{c}}_{r}^{(4)}(t;0)$ becomes

\begin{equation}
\begin{array}{ll}
\mathbb{\mathfrak{c}}_{r}^{(4)}(t;0)= & \overset{\infty}{\underset{b=-\infty}{\sum}}\overset{M-1}{\underset{\underset{\textrm{odd}}{m=0}}{\sum}}\textrm{E}\left[\tilde{x}_{m}^{b}\left(\tilde{x}_{m}^{b}\right)^{*}\right]\\ & \hspace{-0.6cm}\times g(t-(m+L)T-b(M+L)T)\\
 & \hspace{-0.6cm} \times g^{*}(t-(m+L)T-b(M+L)T).\end{array}\label{eq:14}\end{equation}
 
   By considering that  $m=2n+1$, $n=0,...,N-1$, and
 by replacing (\ref{eq:1}) into (\ref{eq:14}) with $Q=2$ $^{3}$, $\mathbb{\mathfrak{c}}_{r}^{(4)}(t;\tau)$ becomes 
 
 \begin{equation}
\begin{array}{l}
\mathbb{\mathfrak{c}}_{r}^{(4)}(t;0)=\overset{\infty}{\underset{b=-\infty}{\sum}}\overset{M-1}{\underset{\underset{\textrm{odd}}{m=0}}{\sum}}\left(\frac{1}{2N}\left(1-e^{j\pi}\right)\right)^{2}\\\hspace{1cm}\times\textrm{E}\left[\overset{N-1}{\underset{p=0}{\sum}}\frac{x_{p}^{b}(x_{p}^{b})^{*}}{(1-e^{j2\pi(\frac{\left(n-p\right)}{N}+\frac{1}{2N})})(1-e^{-j2\pi(\frac{\left(n-p\right)}{N}+\frac{1}{2N})})}\right]\\
\hspace{1cm}\times g(t-(m+L)T-b(M+L)T)\\
\hspace{1cm} \times g^{*}(t-(m+L)T-b(M+L)T))\\
\qquad\quad\quad=\frac{c_{x}}{2N^{2}}\overset{\infty}{\underset{b=-\infty}{\sum}}\overset{M-1}{\underset{\underset{\textrm{odd}}{m=0}}{\sum}}\left[\overset{N-1}{\underset{p=0}{\sum}}\frac{1}{1-\textrm{cos}(\pi(\frac{2n-2p+1}{N}))}\right]\\
\hspace{1cm} \times g(t-(m+L)T-b(M+L)T)\\  \hspace{1cm} \times g^{*}(t-(m+L)T-b(M+L)T).\end{array}\label{eq:15}\end{equation}
By evaluating the summation $\overset{N-1}{\underset{p=0}{\sum}}\frac{1}{1-\textrm{cos}(\pi(\frac{2n-2p+1}{N}))}$
 numerically, one finds that it equals ${N^{2}}$/2. Therefore, (\ref{eq:15}) becomes

\begin{equation}
\begin{array}{l}
 \hspace{-0.15cm}\mathbb{\mathfrak{c}}_{r}^{(4)}(t;0)=\frac{c_{x}}{4} \displaystyle \hspace{-0.25cm} \sum_{b=-\infty}^{\infty}  \hspace{-0.02cm} \sum_{\underset{\textrm{odd}}{m=0}}^{M-1}g(t-(m+L)T-b(M+L)T)\\ \hspace{1.3cm}\times g^{*}(t-(m+L)T-b(M+L)T).
\end{array}
\label{eq:16}\end{equation}
With $m=M-L+u$,  using (\ref{eq:3}), and following the same procedure as for $\mathbb{\mathfrak{c}}_{r}^{(3)}(t;0)$ and $\mathbb{\mathfrak{c}}_{r}^{(4)}(t;0)$, one can easily show that  

\vspace{-0.4cm} 
 \begin{equation}
 \begin{array}{l}
 \hspace{-0.15cm}\mathbb{\mathfrak{c}}_{r}^{(1)}(t;0)=\frac{c_{x}}{4} \displaystyle \hspace{-0.25cm} \sum_{b=-\infty}^{\infty}\sum_{\underset{\textrm{odd}}{m=M-L}}^{M-1}g(t-(m-M+L)T \\ -b(M+L)T)g^{*}(t-(m-M+L)T-b(M+L)T).
 \end{array}
\label{eq:19}\end{equation}
and
\begin{equation}
\begin{array}{l}
 \hspace{-0.15cm}\mathbb{\mathfrak{c}}_{r}^{(2)}(t;0)=\frac{c_{x}}{4} \displaystyle \hspace{-0.25cm} \sum_{b=-\infty}^{\infty}\sum_{\underset{\textrm{odd}}{m=M-L}}^{M-1}g(t-(m-M+L)T \\ -b(M+L)T)g^{*}(t-(m-M+L)T-b(M+L)T).
\end{array}
\label{eq:20}\end{equation}
Finally, by substituting (\ref{eq:13}), (\ref{eq:16}), (\ref{eq:19}), and (\ref{eq:20}), and using $u=m-M+L$ in  the expressions of $\mathbb{\mathfrak{c}}_{r}^{(1)}(t;\tau)$ and
$\mathbb{\mathfrak{c}}_{r}^{(2)}(t;\tau)$ and $u=m+L$ in the expressions
of $\mathbb{\mathfrak{c}}_{r}^{(3)}(t;\tau)$ and $\mathbb{\mathfrak{c}}_{r}^{(4)}(t;\tau)$,
(\ref{eq:11}) becomes

 \begin{equation}
 \begin{array}{l} 
c_{r}(t;0)=\frac{c_{x}}{4} \displaystyle  \sum_{b=-\infty}^{\infty}\sum_{u=0}^{M+L-1}g(t-uT-b(M+L)T) \\
\hspace{1.15cm} \times g^{*}(t-uT-b(M+L)T).\end{array} \label{eq:21}\end{equation}

With $k=u+b(M+L)$, $c_{r}(t;0)$ can be further written as

\begin{equation}
c_{r}(t;0)=  \frac{c_{x}}{4}\left[g(t)g^{*}(t)\right]\otimes\overset{\infty}{\underset{k=-\infty}{\sum}}\delta(t-kT).\label{eq:22}\end{equation}

\end{IEEEproof}
By taking the Fourier transform of (\ref{eq:44}) with respect to $t$, one can easily obtain the CFs, and further using (\ref{eq:4}), the CAF at CF $\beta$ and delay $\tau$ can be expressed as

\vspace{-0.25cm}
 \begin{subequations}\label{eq:46}
\begin{multline} 
\bullet \textrm{ for }{\tau}=\pm{\tau}_{s},\:0\leq{\tau}_{s}<T, \textrm{ and } {\beta}=kT^{-1},\, k\textrm{ integer}, \textcolor{white}{............................................................................................} \\
 c_{r}({\beta};{\tau})=\left(\frac{T-{\tau}_{s}}{4T}+\frac{{\tau}_{s}}{2NT}\frac{1}{1-e^{j\pi\frac{1}{N}}}\right)c_{x}{T^{-1}} \\ \hspace{1.3cm} \times \int_{-\infty}^{\infty}\left[g(t)g^{*}(t)\right]e^{-j2\pi{\beta}t}dt, \textcolor{white}{.................} 
 \end{multline} 
 
 \vspace{-0.25cm}
\begin{multline}
\bullet \textrm{ for }{\tau}\hspace{-0.1cm} = \hspace{-0.1cm} {\mu}T+\textrm{sgn}(\mu){\tau}_{s},\: \mu=\pm1,\pm3,\ldots, \\ \pm(M-L-1),\:0\leq{\tau}_{s}<T, \textrm{ and } {\beta}=kT^{-1},\, k\textrm{ integer},  \\
c_{r}({\beta};{\tau})=\frac{T-{\tau}_{s}}{T}A(\mu){T^{-1}} \hspace{-0.1cm} \int_{-\infty}^{\infty}\hspace{-0.1cm} \left[g(t)g^{*}(t)\right]e^{-j2\pi{\beta}t}dt,
\end{multline} 

\vspace{-0.25cm}
\begin{multline}
\bullet \textrm{ for }{\tau} ={\mu}T+\textrm{sgn}(\mu){\tau}_{s}, \: \mu=\pm2,\pm4,\ldots,\\ \pm(M-L),\:0\leq{\tau}_{s}<T,\textrm{ and } {\beta}=kT^{-1},\, k\textrm{ integer},  \\
c_{r}({\beta};{\tau})\hspace{-0.1cm} = \hspace{-0.1cm} \frac{{\tau}_{s}}{T}A(\mu+1){T^{-1}}\int_{-\infty}^{\infty} \hspace{-0.1cm} \left[g(t)g^{*}(t)\right]e^{-j2\pi{\beta}t}dt,
 \end{multline} 

\vspace{-0.25cm}
\begin{multline}
\bullet \textrm{ for }{\tau}={\mu}T+\textrm{sgn}(\mu){\tau}_{s},\:\mu=\pm M,\:0\leq {\tau}_{s}<T,  \\ \textrm{ and } {\beta}=b\left[(M+L)T \right]^{-1},\, b\textrm{ integer}, 
 \\
c_{r}({\beta};{\tau})=(\frac{T-{\tau}_{s}}{4T}+\frac{{\tau}_{s}}{2NT}\frac{1}{1-e^{\pm j\pi\frac{1}{N}}})c_{x}{\left[(M+L)T\right]^{-1}} \\ \hspace{1.3cm} \times \int_{-\infty}^{\infty} \overset{L-1}{\underset{u=0}{\sum}}\left[g(t-uT)g^{*}(t-uT)\right]e^{-j2\pi{\beta}t}dt,
\end{multline}

\vspace{-0.25cm}
\begin{multline}
\bullet \textrm{ for }{\tau}={\mu}T+\textrm{sgn}(\mu){\tau}_{s},\:\mu=\pm(M-L+1), \\ \pm(M-L+3), \ldots,\pm(M+L-1),0\leq {\tau}_{s}<T,\\ \textrm{ and } {\beta}=b\left[(M+L)T \right]^{-1},\, b\textrm{ integer}, \textcolor{white}{.....................................................................................} 
\\
c_{r}({\beta};{\tau})=\frac{T-{\tau}_{s}}{T}A(\mu){\left[(M+L)T\right]^{-1}}  \textcolor{white}{.........................................................................................} 
\\ \hspace{1.3cm} \times \int_{-\infty}^{\infty} \overset{L-1}{\underset{u=0}{\sum}}\left[g(t-uT)g^{*}(t-uT)\right]e^{-j2\pi{\beta}t}dt, \end{multline}

\vspace{-0.25cm}
\begin{multline}
\bullet \textrm{ for }{\tau}={\mu}T+\textrm{sgn}(\mu){\tau}_{s},\:\mu=\pm(M-L+2),\ldots, \\ \pm(M-2),\pm(M+2),\ldots\pm(M+L-2), 0\leq {\tau}_{s}<T,\\ \textrm{ and } {\beta}=b\left[(M+L)T \right]^{-1},\, b\textrm{ integer}, \textcolor{white}{............................................................................} \\
c_{r}({\beta};{\tau})=\frac{{\tau}_{s}}{T}A(\mu+1){\left[(M+L)T\right]^{-1}} \\ \hspace{1.2cm} \times\int_{-\infty}^{\infty} \overset{L-1}{\underset{u=0}{\sum}}\left[g(t-uT)g^{*}(t-uT)\right]e^{-j2\pi{\beta}t}dt,\end{multline}

\vspace{-0.25cm}
\begin{multline}  \bullet \textrm{ otherwise } \\
c_{r}({\beta};{\tau})=0. \textcolor{white}{................................................}
 \end{multline}
\end{subequations}

The analytical closed-form expression for the CAF of the discrete-time SC-FDMA signals $r(n)=$ $r(t)|$ $_{t=nf_{s}}$, $f_{s}=\rho/T$,
with $\rho$ as the oversampling factor, can be straightforwardly obtained from (\ref{eq:46}) by replacing $\beta$ with $\tilde{\beta}f_{s}$ and $\tau$ with $\tilde{\tau}/f_{s}$, according to  (\ref{eq:6}). Note that this expression is not provided here due to space considerations.

\begin{figure*}
	\centering
	\subfigure[]{%
	\label{fig:a}%
		\includegraphics[width=0.52\textwidth]{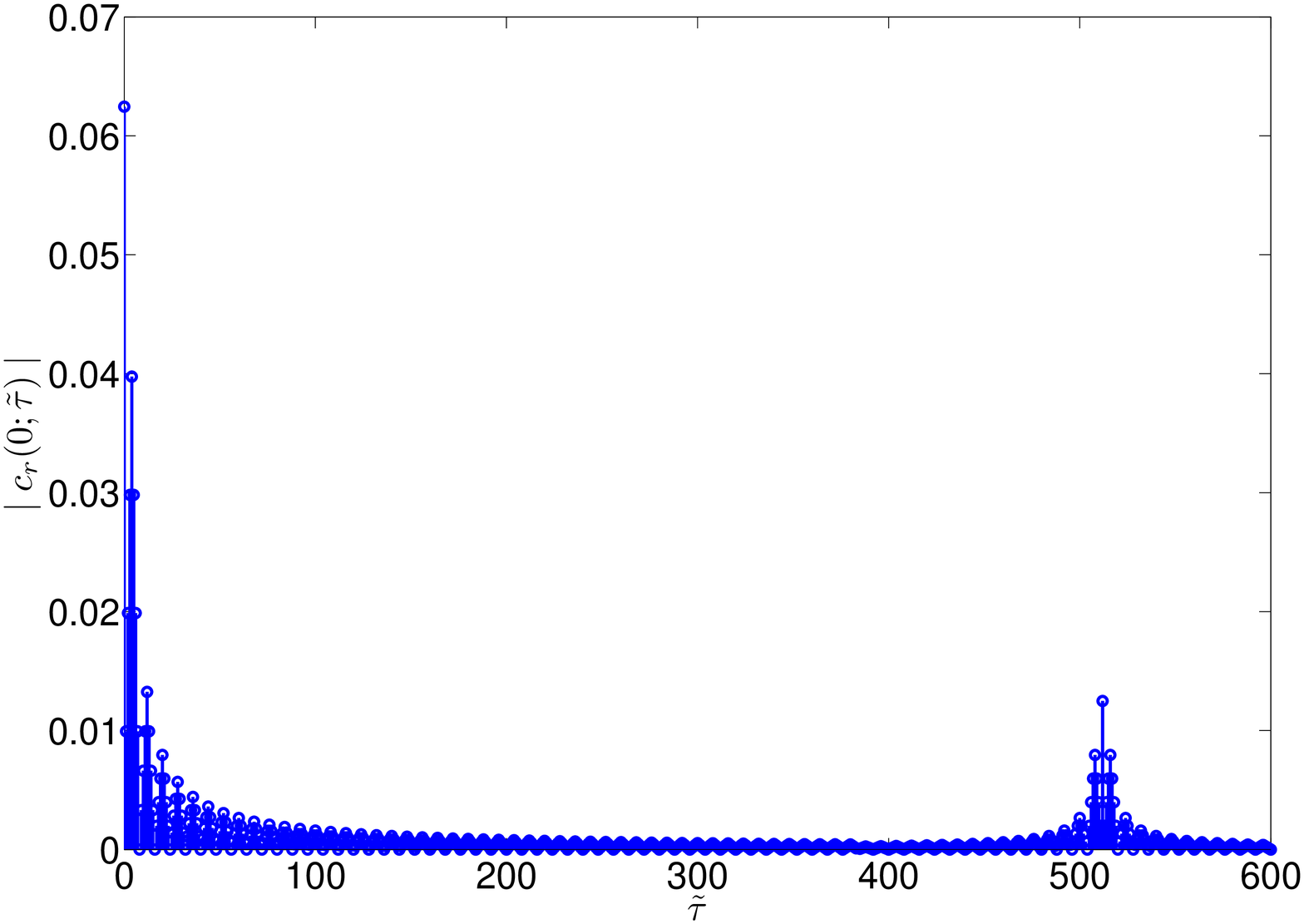}
}%
	\subfigure[]{%
	\label{fig:b}%
		\includegraphics[width=0.52\textwidth]{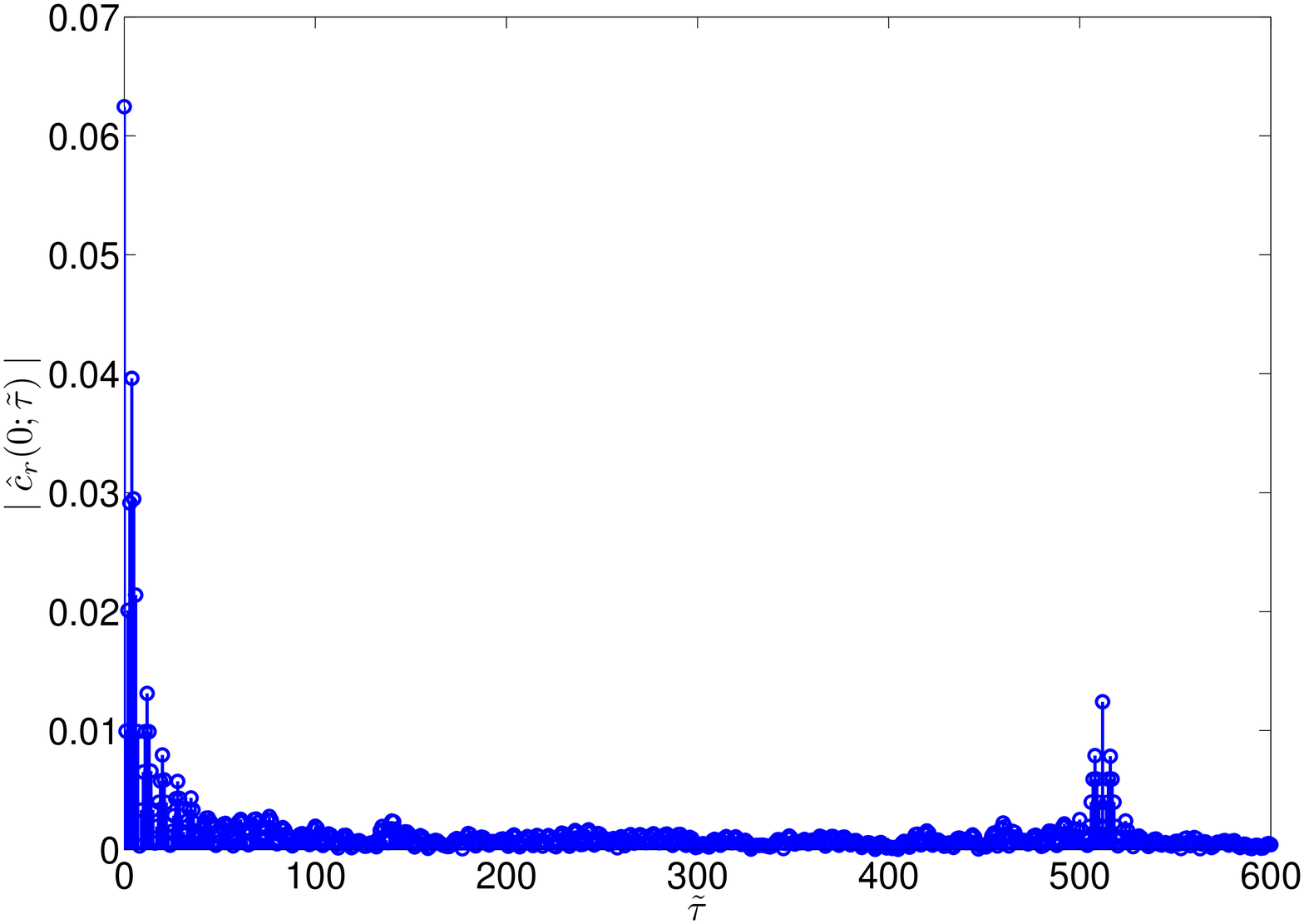}
}%
	\caption{The CAF magnitude at zero CF ($\tilde{\beta}=0$)
versus positive delays, $\tilde{\tau}$: (a) theoretical and (b) simulation results.} 

	\label{fig:CP1}%
\end{figure*}

\begin{figure*}
	\centering
	\subfigure[]{%
	\label{fig:a}%
		\includegraphics[width=0.52\textwidth]{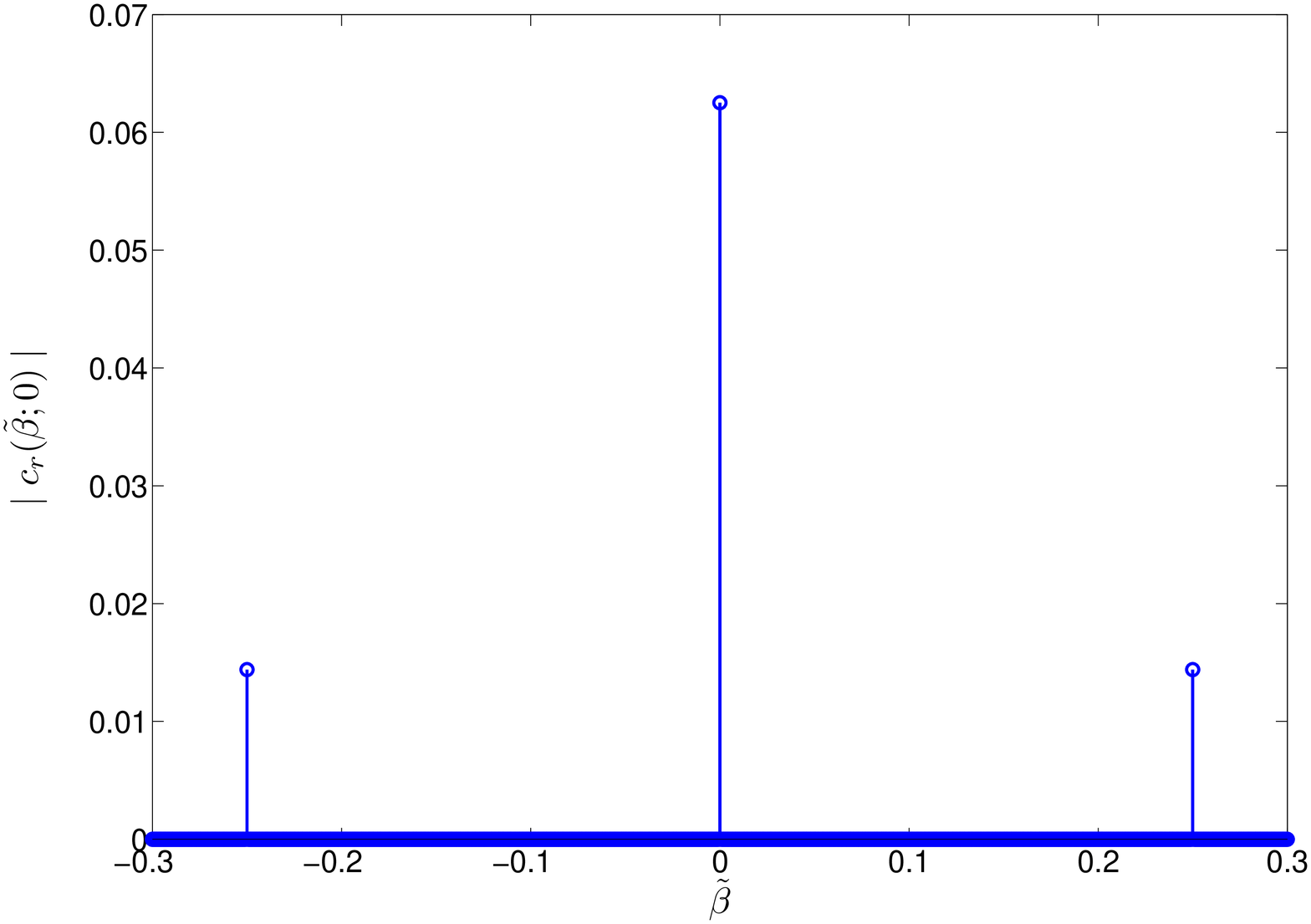}
}%
	\subfigure[]{%
	\label{fig:b}%
		\includegraphics[width=0.52\textwidth]{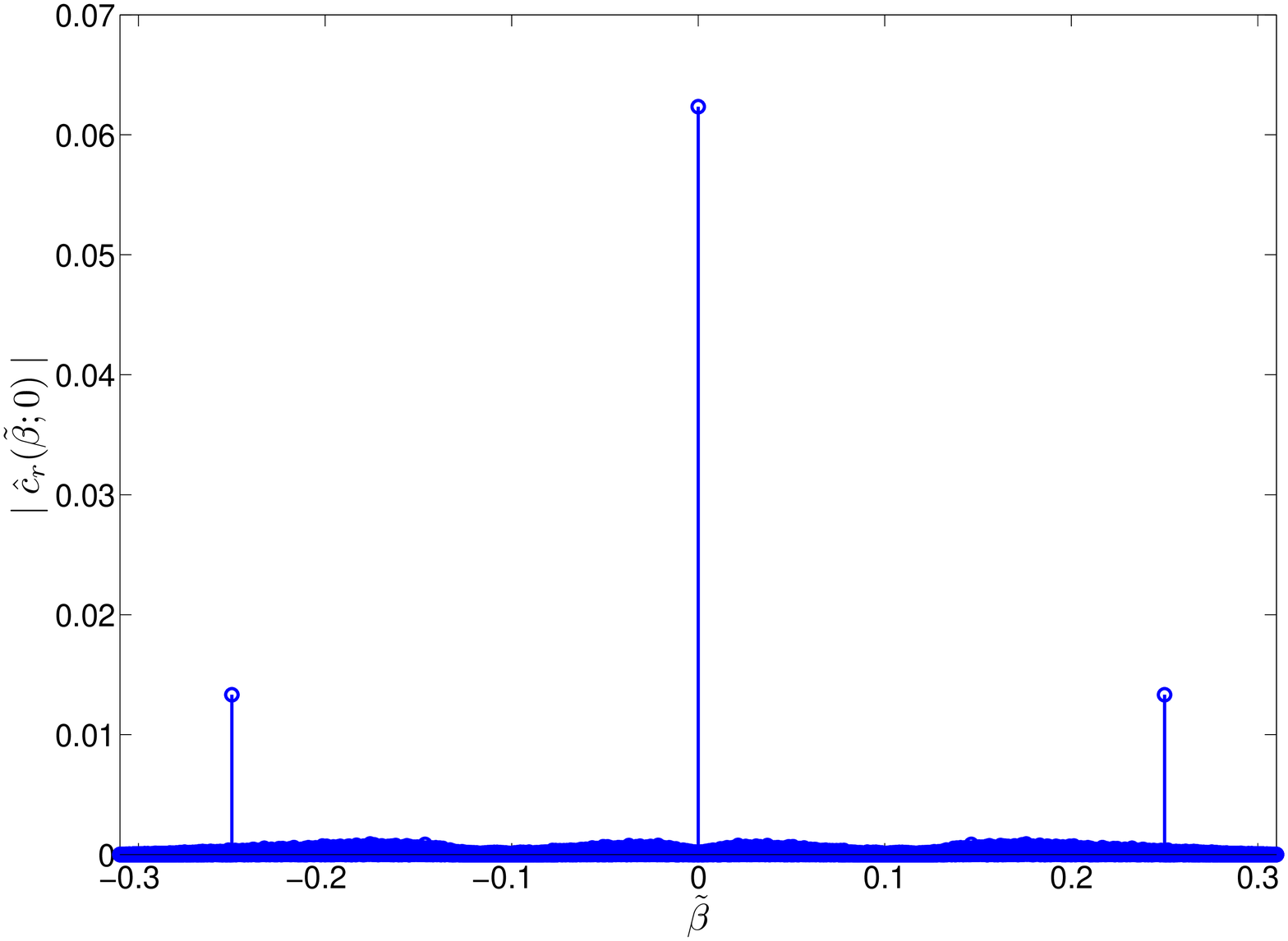}
}%
	\caption{ The CAF magnitude at zero delay ($\tilde{\tau}=0$) versus CF, $\tilde{\beta}$: (a) theoretical and (b) simulation results.} 

	\label{fig:CF1}%
\end{figure*}

In the following, theoretical and simulation results are presented  for the CAF magnitude of the LTE SC-FDMA signals in the absence of noise for different delays and CFs. The parameters  of the LTE SC-FDMA signals are set as in Section V. A, with long CP, and the observation time is 20 ms. Fig. \ref{fig:CP1}  shows the magnitude of the theoretical and estimated CAF magnitude at zero CF versus positive delays. 
Additionally, Fig. \ref{fig:CF1} depicts the CAF magnitude at zero delay versus CFs.  
From these results, one can easily see that the theoretical findings are in agreement with the simulation results.  Note that the non-zero CAF values obtained from simulations, which are theoretically zero, are due to a finite observation time; however, they are not statistically significant.

\section{Proposed Algorithm for the Detection of the LTE SC-FDMA Signals} 

Here, results obtained in Section III for the CAF of the LTE SC-FDMA signals are exploited to develop a cyclostationarity-based algorithm for their detection. We first introduce the SC-FDMA signal features, and then describe the test used for decision-making and the proposed algorithm, as well as discuss its computational complexity.

\vspace{-0.1cm}
\subsection{ Signal Feature Used for Detection}
The  SC-FDMA signals are detected in the frequency bands allocated to the LTE system \cite{e-utra}; accordingly, the values of $M$ and $N$ are known. The signal is down-converted and oversampled, and the baseband discrete-time signal, $r(u),\,0\leq u\leq U_{s}-1$, is exploited for detection.
Based on the theoretical results presented in Section III, one can see that  the CAF magnitude of the received LTE SC-FDMA has the following properties:
\begin{itemize}
\item It has non-zero values at zero  CF and for delays around $\pm\rho M$, with the latter due to the existence of the CP.
\item It is non-zero at CFs equal to $\pm\rho^{-1}$ and for delays around zero. 
\end{itemize}
These particular properties are exploited for the LTE SC-FDMA signal detection. Under hypothesis $\mathcal{H}_{1}$, we assume that the LTE SC-FDMA is present, while under $\mathcal{H}_{0}$
it is not.

\vspace{-0.1cm}

\subsection{ Cyclostationarity Test Used for Decision-Making}

Based on the underlying theory of the cyclostationarity test introduced
in \cite{Cyclo_test}, which verifies if CAF has a CF at $\tilde{\beta}$
for delay $\tilde{\tau}$,  a test for two CFs ($\tilde{\beta_{1}}=0$
and $\tilde{\beta_{2}}=\rho^{-1}$) and two delays ($\tilde{\tau}_{1}=\rho M$
and $\tilde{\tau}_{2}=0$) is developed, such that  the above mentioned
properties of the LTE SC-FDMA signals are exploited. 

\begin{flushleft}
The test used for decision-making is as follows: 
\par\end{flushleft}
\begin{itemize}
\item The CAF of the received signal $r(u)$ is estimated (from $U_{s}$
samples) at each tested frequency $\tilde{\beta_{i}}$ and delay $\tilde{\tau}_{i}$,
$i=1,\,2,$ and a vector $\hat{c}_{i}$ is formed as 
\end{itemize}
\begin{equation}
\hat{c}_{i}=[\textrm{Re}\{\hat{c}_{r}(\tilde{\beta_{i}};\tilde{\tau}_{i})\}\,\textrm{Im}\{\hat{c}_{r}(\tilde{\beta}_{i};\tilde{\tau}_{i})\}],\label{eq:1-2}\end{equation}
where $\textrm{Re }\{.\}$ and $\textrm{Im }\{.\}$ are the real and
the imaginary parts, respectively.

\begin{itemize}
\item A statistic $\Psi_{i}$ , $i=1,\,2,$ is computed for each tested frequency
$\tilde{\beta}_{i}$ and delay $\tilde{\tau}_{i}$ as \cite{Cyclo_test}
\end{itemize}
\begin{equation}
\Psi_{i}=U_{s}\hat{c}_{i}\hat{\sum}_{i}^{-1}\hat{c}_{i}^{\intercal},\label{eq:2-3}\end{equation}
where the superscripts -1 and $\intercal$ denote the matrix inverse
and transpose, respectively, and $\hat{\sum}_{i}$ %
\footnote{Note that although $Q_{2,0}$ and $Q_{2,1}$ depend on $i$, this
dependency was not shown for simplicity of notation. %
} is an estimate of the covariance matrix

\vspace{-0.2cm}
\begin{equation}
 {\sum}_{i} = \left[\begin{array}{cc}
\textrm{Re}\left\{ \frac{\left(Q_{2,0}+Q_{2,1}\right)}{2}\right\}  & \textrm{Im} \left\{ \frac{\left(Q_{2,0}-Q_{2,1}\right)}{2}\right\} \\
\textrm{Im}\left\{ \frac{\left(Q_{2,0}+Q_{2,1}\right)}{2}\right\}  & \textrm{Re}\left\{ \frac{\left(Q_{2,1}-Q_{2,0}\right)}{2}\right\} \end{array}\right],\label{eq:3-2}\end{equation}
with 
\begin{equation}
Q_{2,0}=\underset{U_{s\rightarrow\infty}}{\textrm{lim}}U_{s}\textrm{Cum}[\hat{c}_{r}(\tilde{\beta_{i}};\tilde{\tau}_{i}),\hat{c}_{r}(\tilde{\beta_{i}};\tilde{\tau}_{i})],\label{eq:4-1}\end{equation}
and 
\begin{equation}
Q_{2,1}=\underset{U_{s\rightarrow\infty}}{\textrm{lim}}U_{s}\textrm{Cum}[\hat{c}_{r}(\tilde{\beta_{i}};\tilde{\tau}_{i}),\hat{c}_{r}^{*}(\tilde{\beta_{i}};\tilde{\tau}_{i})].\label{eq:5}\end{equation}

For a zero-mean process, the covariances $Q_{2,0}$ and $Q_{2,1}$
are given respectively as \cite{Cyclo_test}

\vspace{-0.2cm}
\begin{equation}
\begin{array}{l}
\hspace{-0.25cm}Q_{2,0}=\underset{U_{s\rightarrow\infty}}{lim}U_{s}^{-1} \displaystyle \sum_{l=0}^{U_{s}-1}\sum_{\xi=-\infty}^{\infty} \hspace{-0.15cm} \textrm{Cum}\left[f\left(l;\tilde{\tau}_{i}\right),f\left(l+\xi;\tilde{\tau}_{i}\right)\right] \\ \hspace{4.5cm} \times e^{-j2\pi2\tilde{\beta_{i}}l}e^{-j2\pi\tilde{\beta_{i}}\xi},\end{array} \label{eq:6-1}\end{equation}
and
\begin{equation}
\begin{array}{l}
\hspace{-0.25cm}Q_{2,1}=\underset{U_{s\rightarrow\infty}}{lim}U_{s}^{-1} \displaystyle \sum_{l=0}^{U_{s}-1}\sum_{\xi=-\infty}^{\infty} \hspace{-0.15cm}\textrm{Cum}\left[f\left(l;\tilde{\tau}_{i}\right),f^{*}\left(l+\xi;\tilde{\tau}_{i}\right)\right] \\ \hspace{4.5cm} \times e^{-j2\pi(-\tilde{\beta_{i}})\xi l},\end{array} \label{eq:7-1}\end{equation}
where $f\left(l;\tilde{\tau}_{i}\right)=r(l)r^{*}(l+\tilde{\tau}_{i})$
is the second-order (one-conjugate) lag product.

Moreover, the estimators of the covariance $Q_{2,0}$ and $Q_{2,1}$ are given
respectively by \cite{Cyclo_test}


\vspace{-0.2cm}
\begin{equation}
\begin{array}{l}
\hat{Q}_{2,0}=(U_{s}U_{sw})^{-1} \displaystyle \sum_{s=-(U_{sw}-1)/2}^{(U_{sw}-1)/2}W(s)F_{\tilde{\tau}_{i}}(\tilde{\beta_{i}}-sU_{s}^{-1}) \\ \hspace{4.4cm}\times F_{\tilde{\tau}_{i}}(\tilde{\beta}_{i}+sU_{s}^{-1}), \end{array} \label{eq:6-1}\end{equation}
and
\begin{equation}
\begin{array}{l}

\hat{Q}_{2,1}=(U_{s}U_{sw})^{-1} \displaystyle  \sum_{s=-(U_{sw}-1)/2}^{(U_{sw}-1)/2}W(s)F_{\tilde{\tau}_{i}}^{*}(\tilde{\beta}_{i}+sU_{s}^{-1}) \\ \hspace{4.4cm}\times F_{\tilde{\tau}_{i}}(\tilde{\beta}_{i}+sU_{s}^{-1}),\end{array} \label{eq:7-1-1}\end{equation}

where $W(s)$ is a spectral window of length $U_{sw}$ and $F_{\tilde{\tau}_{i}}(\tilde{\beta}_{i})=\overset{U_{s}-1}{\underset{u=0}{\sum}}r(u)r^{*}(u-\tilde{\tau}_{i})e^{-j2\pi\tilde{\beta}_{i}u}$.
\begin{itemize}
\item With the test statistics $\Psi_{1}$ and $\Psi_{2}$ calculated based
on the estimated CAF at $\tilde{\beta_{1}}=0$ and $\tilde{\tau}_{1}$=$\rho M$
and at $\tilde{\beta_{2}}=\rho^{-1}$ and $\tilde{\tau}_{2}=$0, respectively,
we form a new test statistic, $\Upsilon=\Psi_{1}+\Psi_{2}$. For
decision-making, we compare $\Upsilon$ against a threshold, $\Gamma$.
If $\Upsilon\geq\Gamma,$ we decide that the LTE SC-FDMA is present
(hypothesis $\mathcal{H}_{1}$); otherwise,  that it is not
(hypothesis $\mathcal{H}_{0}$). By using that the statistics
$\Psi_{1}$ and $\Psi_{2}$ have asymptotic chi-square distribution
with two degrees of freedom \cite{Cyclo_test}, it is straightforward
to find that $\Upsilon$ asymptotically follows a chi-square distribution
with four degrees of freedom. The threshold $\Gamma$ is obtained
from the tables of this chi-squared distribution for a given value
of the probability of false alarm ($P_{fa}$), i.e., $P_{fa}=\textrm{\ensuremath{P_{r}}}\left\{ \Psi\geq\Gamma\mid H_{0}\right\} $ \cite{abramowitz2012handbook}.
A summary of the proposed detection algorithm is provided below.\end{itemize}
\floatname{algorithm}{}
\begin{algorithm}
\renewcommand{\thealgorithm}{}
\caption{\textbf{Summary of the proposed detection algorithm}}
\begin{algorithmic}[0]
\State \textbf{Input:} The observed samples $r(u)$, $u=0,...,U_{s}-1,$
and $\rho$, $M$, and target $P_{fa}$.
\State - Calculate $\hat{c}_{r}(0,\rho M)$ and $\hat{c}_{r}(\rho^{-1},0)$
using (\ref{eq:7}).
\State - Calculate $\Psi_{1}$ using $\hat{c}_{r}(0,\rho M)$, according to (\ref{eq:2-3}).
\State - Calculate $\Psi_{2}$ using $\hat{c}_{r}(\rho^{-1},0)$, according to (\ref{eq:2-3}).
\State - Calculate $\Upsilon=\Psi_{1}+\Psi_{2}$.
\State - Calculate $\Gamma$ based on $P_{fa}$.
\If {$\Upsilon\geq\Gamma$}
\State - LTE SC-FDMA is present ($\mathcal{H}_{1}$ true).
\Else
\State - LTE SC-FDMA is not present ($\mathcal{H}_{0}$ true).
\EndIf
\end{algorithmic}
\end{algorithm}


\textit{C. Complexity Analysis of the Proposed Detection Algorithm}

The computational complexity of the algorithm is basically determined by the calculation of the test statistic $\Upsilon$, which entitles computation of $\Psi_{1}$ and $\Psi_{2}$. In order to obtain the number of operations required for that, in the following we investigate the complexity of estimating the CAF at $\tilde{\beta}_{i}$ and $\tilde{\tau}_{i}$, $\hat{c}_{r}(\tilde{\beta}_{i},\tilde{\tau}_{i})$, $i=1, 2,$.
As such, according to (\ref{eq:7}), the estimation of CAF at  $\tilde{\beta}_{1}=0$ and $\tilde{\tau}_{1}={\rho}M$ requires $U_s$ complex multiplications and $U_{s}-1$ complex additions, while $2U_s$ complex multiplications and $U_{s}-1$ complex additions are required for $\tilde{\beta}_{2}={\rho}^{-1}$ and $\tilde{\tau}_{2}=0$.
Furthermore, based on (\ref{eq:3-2}),  (\ref{eq:6-1}), (\ref{eq:7-1-1}), and the expression of $\Psi_{i}$ in (\ref{eq:2-3}), one can find that the number of complex multiplications, complex additions, and real operations needed to calculate $\Psi_{i}$ is $(U_{s}/2)log_{2}U_{s}+2U_{sw}$, $U_{s}log_{2}U_{s}+2(U_{sw}-1)$, and  $9U_{sw}+26$, respectively.
Moreover, by considering  $\Psi_{1}$, $\Psi_{2}$, $\Upsilon$, as well as the comparison between $\Gamma$ and $\Upsilon$, and using the fact that a complex multiplication requires 6 floating point operations (flops),  a complex addition requires 2 flops, and a real operation requires 1 flop, the total number of flops  required by the algorithm equals $10U_{s}log_{2}U_{s}+22U_{s}+50U_{sw}+42$ \footnote{Note that the common terms which appeared in the computation of the statistic were counted only once.}. For example, with $U_{s}=$64000 (12.8 ms  observation time) and $U_{sw}=$ $0.006U_{s}$, the proposed algorithm needs 11,645,343 flops, while with $U_{s}=$32000  (6.4 ms  observation time), it requires 5,502,692 flops. Practically speaking, with a microprocessor that can execute up to 79.2 billion flops per second
\footnote{{[}Online{]}. Available: http://ark.intel.com/Product.aspx?id=47932 \\ \&processor=i7-980X\&spec-codes=SLBUZ.},  a decision can be performed in approximately 0.147 ms when the observation time
is 12.8 ms and in 0.069 ms when the observation time is 6.4 ms. It is worth noting that the computational time is significantly lower than the observation time. Also,
there is a tradeoff between complexity and performance, i.e., a longer
observation time will lead to an increased complexity, but also to an improved performance, as it will be shown in Section~V.

\section{Simulation and Experimental Results} 

\subsection{ Simulation Setup} 
The performance of the algorithm proposed for the detection of the
LTE SC-FDMA signals used in the uplink transmission is investigated
here.  Unless otherwise mentioned, the following parameter
values were employed for the SC-FDMA signals \cite{e-utra}: 1.4 MHz bandwidth, $N=72,$ $ M=128,$ $\rho=4$.
The subcarrier spacing was set to $\triangle f=15$ kHz, $L$/$M=$1/4 for
long CP, and $L$/$M$=10/128 for the first symbol in the slot
and $L$/$M= $9/128 for the remaining symbols for short CP.
An RRC with 0.35 roll-off factor was employed at the transmit-side and 16-QAM modulation with unit variance was considered. The impairments which affected the received signals were:
500 kHz carrier frequency offset and uniformly distributed phase and
timing offsets over {[}$-\pi,\pi$) and {[}$0,1$), respectively.
We considered the additive white Gaussian noise (AWGN), and ITU-R
pedestrian and vehicular A channels \cite{molisch2011wireless}. The
maximum Doppler frequencies equal 9.72 Hz and 194.44 Hz for the pedestrian
and vehicular fading channels, respectively. The out-of-band noise
was removed at the receive-side with a 13 order low-pass Butterworth
filter, and the SNR was set at the output of this filter. The probability
of detection, $P_{d}$, is used as a performance measure; this is estimated
based on 1000 Monte Carlo trials. Unless otherwise mentioned, the
sensing times of 6.4 ms and 12.8 ms were used, the probability
of false alarm was equal to $P_{fa}$= 0.01, and a 16-bit uniform quantizer with an overloading factor of 4 was considered for the analog-to-digital converter~(ADC).

\subsection{ Experimental Setup}

\begin{figure}
\begin{centering}
\includegraphics[width=0.5\textwidth]{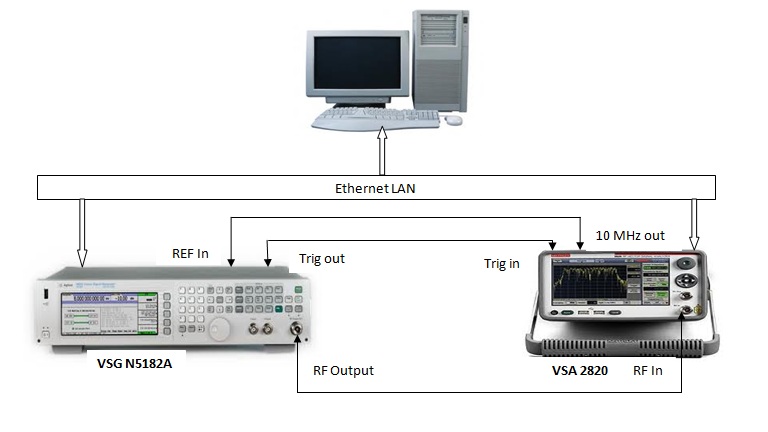}
\par\end{centering}

\caption{Measurement station.\label{fig:setup_hardware}}
\end{figure}

A measurement station was set up as depicted in Fig. \ref{fig:setup_hardware}. This
consists of: 1) a processing and control unit, namely, a personal
computer (PC), 2) an Agilent vector signal generator (VSG N5182A)\footnote{Note that the Agilent VSG N5182A uses a 16-bit quantizer for the ADC \cite{website_Agilent}.},
and 3) a Keithley vector signal analyzer (VSA 2080). The three components
were interfaced via Ethernet. By using the LTE SC-FDMA signal emulation
built into the VSG and the Agilent Studio toolkit, the VSG generated an
RF analog signal, which was transmitted to the VSA through a cable.
The received analog RF signal was down-converted to intermediate frequency,
and then converted to a digital signal, as well as to baseband. Finally,
by using the Keithley SignalMeister, the signal captured with the
VSA was transferred to the PC, where the detection algorithm was applied.
The signal parameters were the same as used in the computer simulations.

\subsection{ Algorithm Performance}

The performance of the proposed algorithm is investigated in terms
of the probability of detection, $P_{d}$, for the LTE SC-FDMA signals, under various conditions.
Results obtained from
both computer simulations (black color) and experiments (red color)
are shown, with a good agreement between simulation and experimental findings.

\begin{figure}
\begin{centering}
\includegraphics[width=0.5\textwidth]{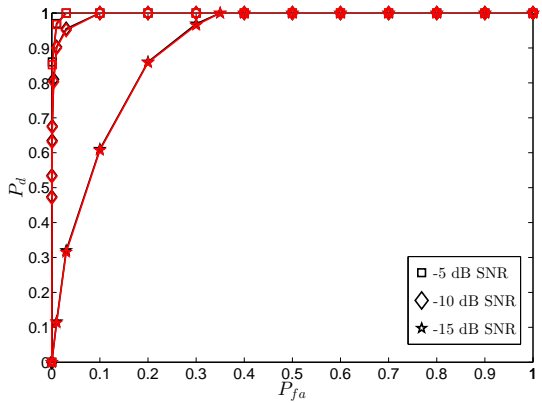}
\par\end{centering}

\caption{The probability of detection, $P_d$, versus $P_{fa}$ for LTE SC-FDMA signals with
long CP affected by pedestrian A channel, for different SNRs and
with 12.8 ms observation time. Simulation (black color) and experimental
(red color) results.\label{fig:falsealarm}}
\end{figure}

 In Fig. \ref{fig:falsealarm}, results for  $P_{d}$ versus $P_{fa}$ are presented at different SNRs, for the ITU-R pedestrian A fading channel and
with 12.8 ms observation time. Clearly, the detection performance
improves with an increase in SNR. For example, with -10 dB SNR, $P_{d}$
approaches 1 at $P_{fa}$ around 0.1, while with -15 dB SNR, this occurs
at $P_{fa}$ around 0.35.

\begin{figure}
\begin{centering}
\includegraphics[width=0.5\textwidth]{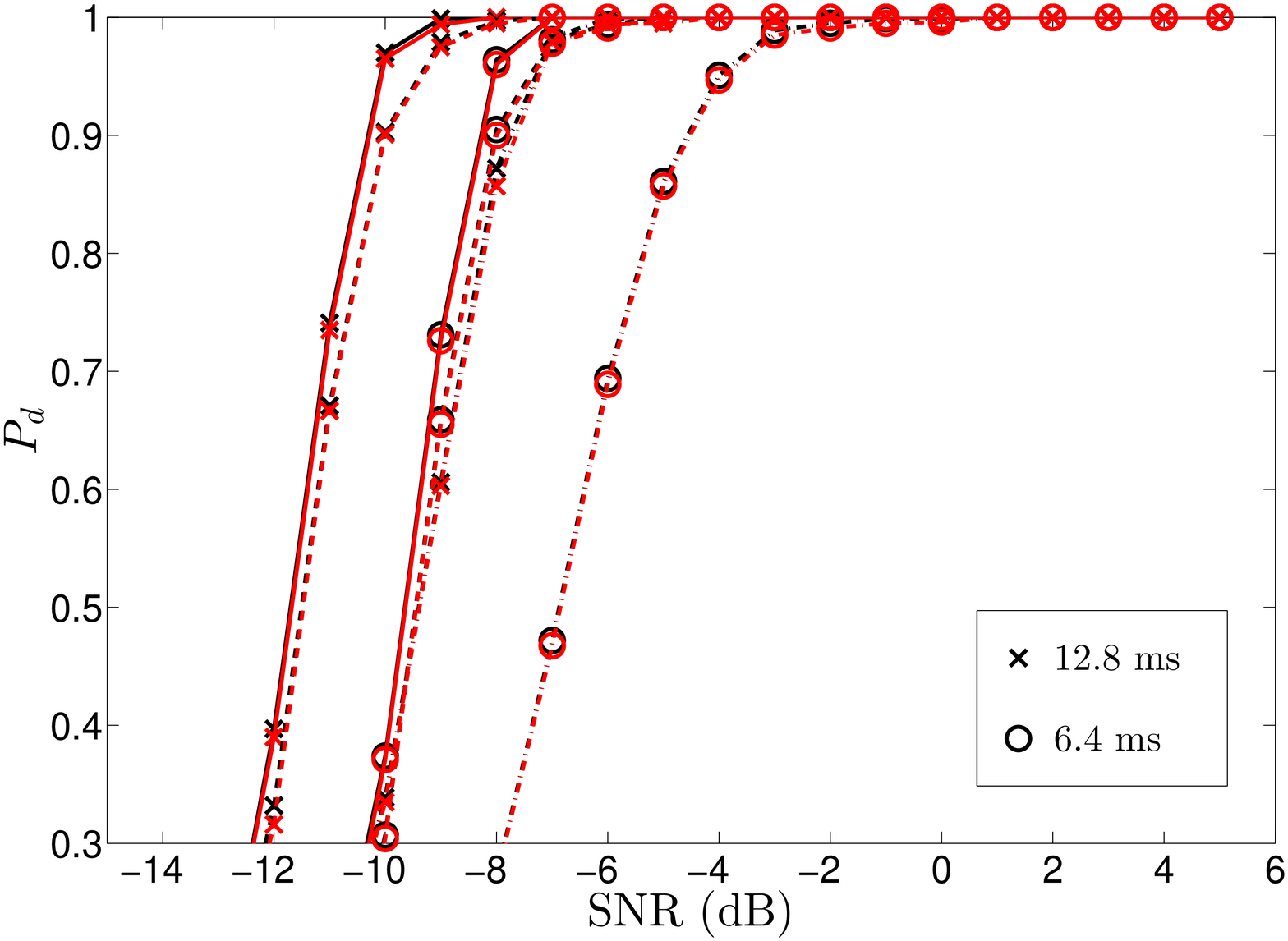}
\par\end{centering}

\caption{The probability of detection, $P_d$, versus SNR for the LTE SC-FDMA signals
with long CP affected by AWGN (solid line), pedestrian A (dashed line),
and vehicular A (dashed-dot line) channels, respectively. Simulation
(black color) and experimental (red color) results.\label{fig:detection}}
\end{figure}

Fig. \ref{fig:detection} shows $P_{d}$ versus SNR for the LTE SC-FDMA signal with long CP, and considering AWGN (solid line), ITU-R pedestrian A (dashed line),
and ITU-R vehicular A fading (dashed-dot line) channels, as well as the observation times of 6.4 ms and 12.8 ms. As expected, the best performance is
obtained in the AWGN channel, followed by the pedestrian and vehicular A channels. While results achieved in the pedestrian A channel are close to those in AWGN, a longer observation time is required to reach the same performance in the vehicular A channel. Also as expected, $P_{d}$ improves as the observation time increases.

\begin{figure}
\begin{centering}
\includegraphics[width=0.5\textwidth]{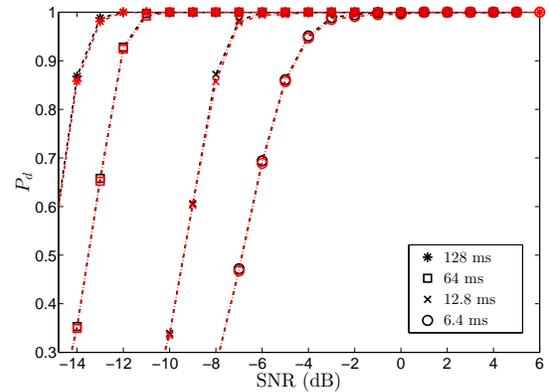}
\par\end{centering}

\caption{The probability of detection, $P_d$, versus SNR for LTE SC-FDMA signals with
long CP affected by vehicular A channel, for different observation
times. Simulation (black color) and experimental (red color) results.\label{fig:sensingtime}}
\end{figure}

Fig. \ref{fig:sensingtime} depicts
$P_{d}$ versus SNR for the vehicular A channel, with different observation times. As previously noticed, the detection performance enhances with an increase in the observation time. For example, while $P_d$ approaches 1 at $-6$ dB SNR with 12.8 ms observation time, such a performance is achieved at $-12$ dB SNR with 128 ms observation time.

\begin{figure}
\begin{centering}
\includegraphics[width=0.5\textwidth]{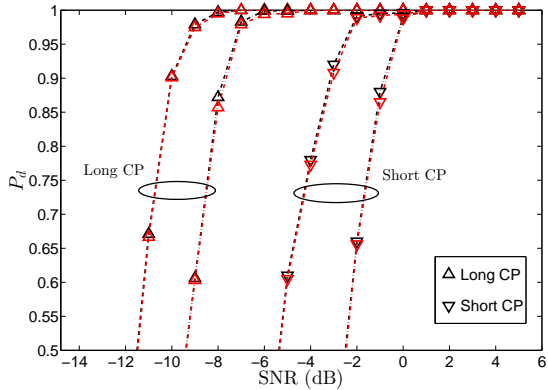}
\par\end{centering}

\caption{The probability of detection, $P_d$, versus SNR for LTE SC-FDMA signals with
long and short CP affected by pedestrian A (dashed line) and
vehicular A (dashed-dot line) channels, with 12.8 ms observation time.
Simulation (black color) and experimental (red color) results.\label{fig:short_long}}
\end{figure}

Fig. \ref{fig:short_long} presents the detection performance for LTE SC-FDMA signals with long and short CPs for the pedestrian and vehicular A channels. As expected, a reduction in the
CP duration adversely affects the performance under the same conditions. This is explained by the reduction in the correlation resulting from the reduced CP duration. 

\begin{figure}
\begin{centering}
\includegraphics[width=0.5\textwidth]{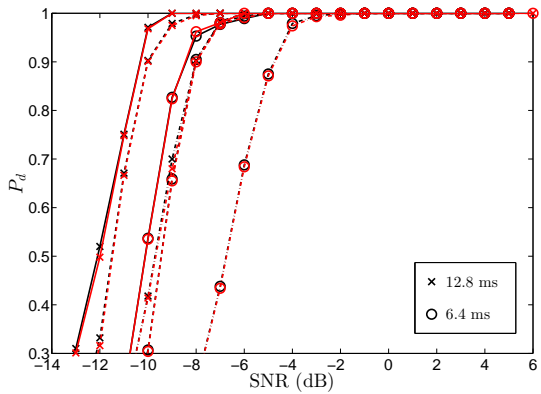}
\par\end{centering}

\caption{The probability of detection, $P_d$, versus SNR for the LTE SC-FDMA signals
with long CP affected by pedestrian A channel, with $\rho=8$ (solid
line), 4 (dashed line), and 2 (dashed-dot line). Simulation (black
color) and experimental (red color) results.\label{fig:rau248}}
\end{figure}

The effect of the oversampling factor, $\rho$, on $P_d$ is shown in Fig. \ref{fig:rau248}; $P_d$ is plotted versus SNR for $\rho=2,4,8$, for the pedestrian A channel, with observation times of 6.4 ms and 12.8 ms.
As expected, the detection performance improves with an increase in $\rho$ for a certain observation time, as the number of samples increase, which in turn leads to more accurate estimates.

\begin{figure}
\begin{centering}
\includegraphics[width=0.5\textwidth]{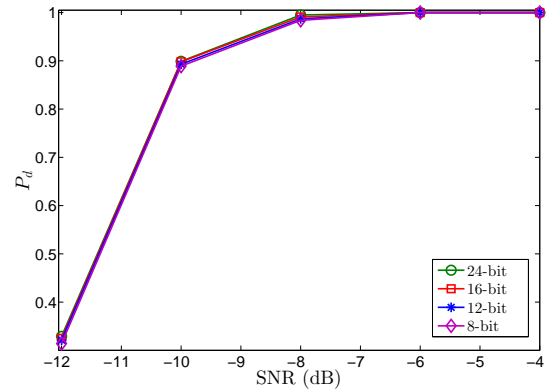}
\par\end{centering}

\caption{The probability of detection, $P_d$, versus SNR for the LTE SC-FDMA signals
with long CP affected by  pedestrian A channel,  with 12.8 ms observation time  and for various numbers of quantization bits.  {\color{white}{.......................................................\hspace{10cm}..................}}\label{fig:Quantization}}
\end{figure}

Fig. \ref{fig:Quantization} shows the effect of the quantization error on the  performance of the proposed detection algorithm when 8, 12, 16, or 24 bits are respectively  used. This error is modeled as an additive noise with uniform distribution over the interval $-\Delta/2$ to $\Delta/2$, where $\Delta$ is the quantization step size which  depends on the bit resolution \cite{gersho1977quantization}. 
 As can be noticed from Fig. \ref{fig:Quantization}, the quantization error does not basically affect the performance. This can be explained, as the proposed algorithm is applied for detecting the presence of LTE SC-FDMA signals rather than detecting the information carried by such signals.

Additionally, we investigated the performance of the proposed detection algorithm in the presence of interference, with results shown in Figs. 12-14. We considered multiuser Gaussian interference, as commonly used in the literature \cite{hayashi2009spectrum,tan2011spectrum,kalathil2013spectrum}.

\begin{figure}
\begin{centering}
\includegraphics[width=0.5\textwidth]{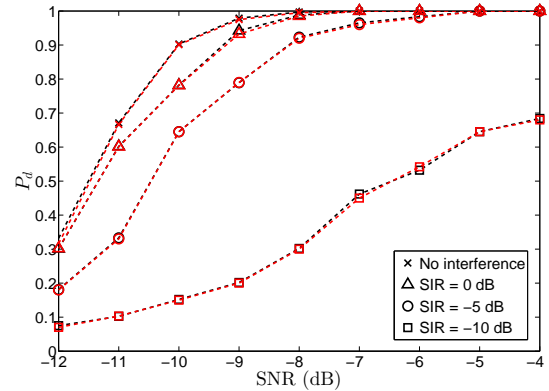}
\par\end{centering}

\caption{The probability of detection, $P_d$, versus SNR for LTE SC-FDMA signals with
long CP affected by pedestrian A channel, with 12.8 ms observation time and for different SIRs. Simulation (black color) and experimental (red color) results.\label{fig:Interfernce}}
\end{figure}

Fig. \ref{fig:Interfernce} depicts $P_d$ versus SNR for the pedestrian A channel, with 12.8 ms observation time and for different values of the signal-to-interference ratio (SIR). Apparently, the detection performance degrades as the SIR reduces. For example, with SIR = 0 dB,  $P_d$ approaches one at around SNR=-7 dB, while the same performance is achieved with SIR=-5 dB at SNR=-5 dB. On the other hand, the detection performance degrades significantly with SIR=-10 dB. 

\begin{figure}
\begin{centering}
\includegraphics[width=0.5\textwidth]{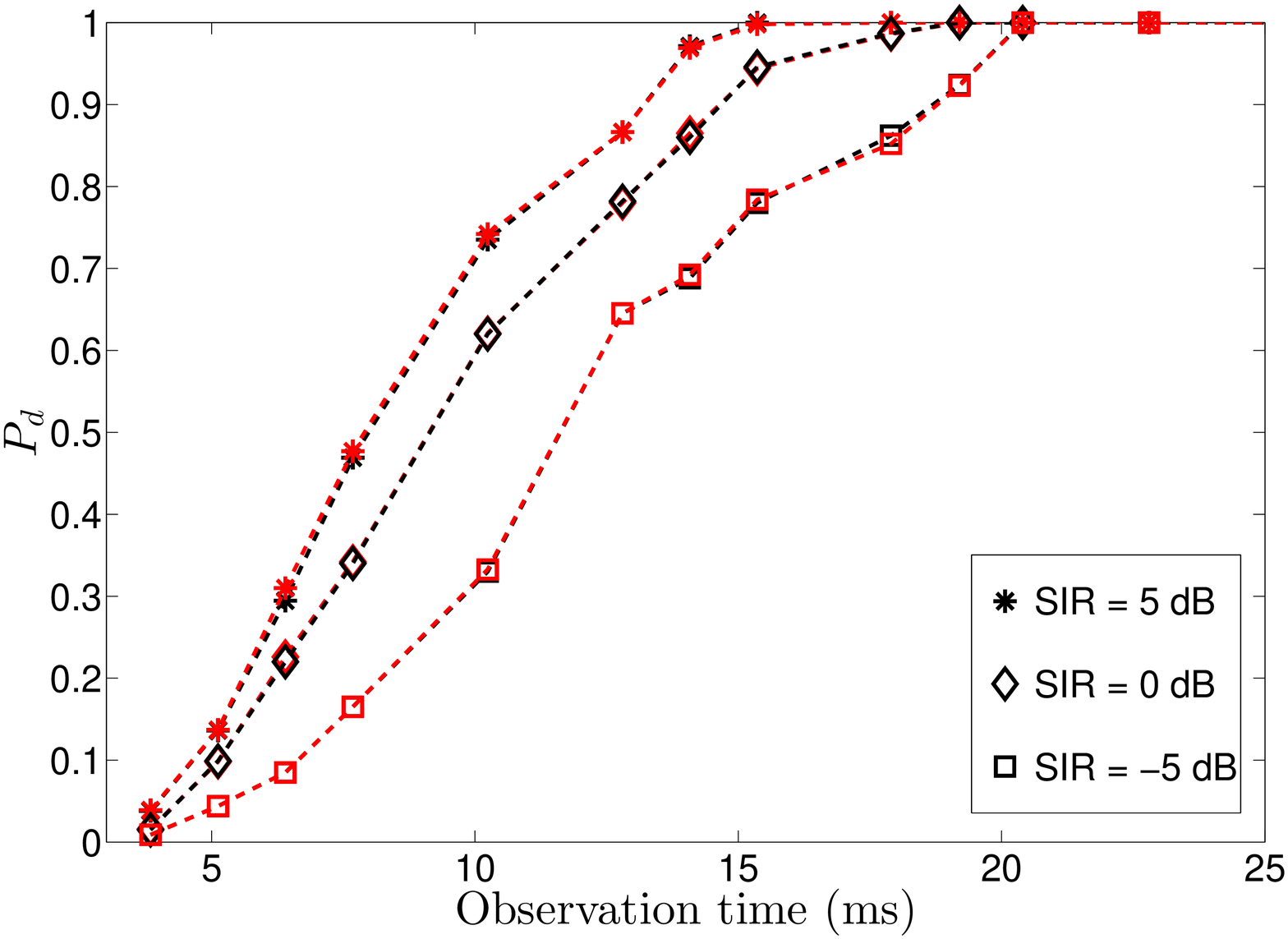}
\par\end{centering}

\caption{The probability of detection, $P_d$, versus the observation time for the LTE SC-FDMA signals
with long CP affected by pedestrian A channel,  for  SNR=-10 dB and  different SIRs.
 Simulation (black color) and experimental (red color) results.\label{fig:Ts_fig}}
\end{figure}

Fig. \ref{fig:Ts_fig} shows   $P_d$ versus the observation time at different SIRs, for the pedestrian A channel, and  SNR= -10 dB.  As expected, the performance improves if the observation interval increases; for example, the detection algorithm fails with SIR=-5 dB and 5 ms observation time, while $P_d$ approaches one with 20 ms observation time for the same SIR. Moreover, less observation time is required to approach $P_d=1$ for higher SIR; for example,
15 ms is needed at SIR=5 dB, while 20 ms is required at SIR=-5 dB.

\begin{figure}
\begin{centering}
\includegraphics[width=0.5\textwidth]{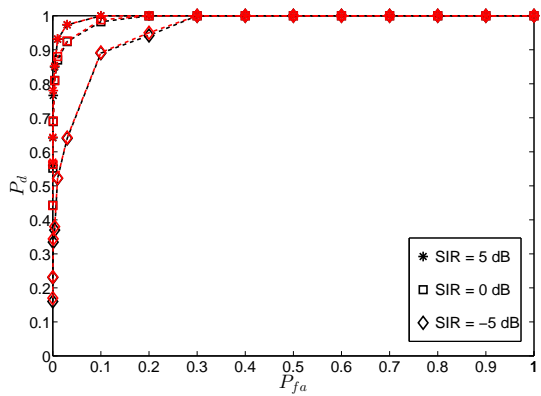}
\par\end{centering}

\caption{The probability of detection, $P_d$, versus $P_{fa}$ for the LTE SC-FDMA signals
with long CP affected by pedestrian A channel, with 12.8 ms observation time, for  SNR=-10 dB and  different SIRs.
 Simulation (black color) and experimental (red color) results.
\label{fig:pfa_SIR}}
\end{figure}

Fig. \ref{fig:pfa_SIR} presents $P_d$ versus $P_{fa}$ at different SIRs, for the pedestrian A channel,  with 12.8 ms observation time, and SNR=-10 dB. As expected,  a better $P_d$ is achieved as $P_{fa}$ increases. For example, $P_d$ is around 0.93 for $P_{fa}=0.01$, while $P_d$ approaches one for $P_{fa}=0.1$ when SIR=5 dB.
Apparently, a better detection performance is  attained when SIR increases. For example, with SIR=5 dB, $P_d$ approaches one at $P_{fa}=0.1$, while the same performance is obtained with  SIR=-5 dB at $P_{fa}=0.3$. 

\section{Conclusion}
In this paper, the  second-order cyclostationarity of the LTE SC-FDMA signal was studied, and closed form expressions for the corresponding CAF and CFs were derived. Furthermore, based on these findings, an algorithm  was developed for the detection of the LTE SC-FDMA signals. Experiments were carried
out using computer simulations and signals generated by laboratory equipment to evaluate the performance of the
proposed algorithm under diverse scenarios, involving various channel conditions, SNRs, SIRs, and observation times. Results showed that this provides a good performance  at low SNRs and with a relatively
short observation time, even in the presence of interference. In addition, it  requires neither frequency and timing synchronization nor channel estimation. The algorithm can be implemented in real world scenarios, 
with reduced complexity.
%
%

\bibliographystyle{IEEEtran}
\bibliography{IEEEabrv,Ref1}

\end{document}